\documentclass[3p]{elsarticle}
\usepackage{etoolbox}
\usepackage{amsmath,amssymb}
\usepackage{hyperref,mathtools,subfigure}
\usepackage{subfigure}
\journal{Materials Transactions}
\usepackage{bm}
\usepackage{multirow}
\usepackage{color}
\usepackage{soul,tabularx}

\DeclareUnicodeCharacter{2212}{-}
\DeclareUnicodeCharacter{2009}{-}
\DeclareUnicodeCharacter{03BA}{-}
\begin{document}

\begin{frontmatter}

\title{Quantum theory of the effect of increasing weak electromagnetic wave by a strong laser radiation in 2D Graphene}
%author group
\author[1]{Tran Anh Tuan}
\author[1]{Nguyen Dinh Nam\corref{cor1}} 
%\ead{nguyendinhnam@hus.edu.vn}
\author[1]{Nguyen Thi Thanh Nhan}
\author[1]{Nguyen Quang Bau\corref{cor2}}
%\ead{nguyenquangbau54@gmail.com, nguyenquangbau@hus.edu.vn}
\cortext[cor1]{nguyendinhnam@hus.edu.vn}
\cortext[cor2]{nguyenquangbau54@gmail.com, nguyenquangbau@hus.edu.vn}
\address[1]{Department of Theoretical Physics, VNU  University of Science, Vietnam National University, Hanoi 100000, Vietnam}

%abstract
\begin{abstract}
Analytic expressions for the absorption coefficient (AC) of a weak electromagnetic wave (EMW) in 2D Graphene under influence of  strong laser radiation are calculated using the quantum kinetic equation (QKE) in the case of electron-optical phonon scattering  in both the absence and presence of a magnetic field perpendicular to the graphene sheet. The dependence of the AC on the intensity $E_{02}$ and the frequency $\Omega_2$ of a weak EMW, on the intensity $E_{01}$ and the frequency $\Omega_1$ of a strong laser radiation, on the temperature T of the system is obtained.  These results are investigated from low temperature to high temperature.  These results are obtained from the QKE method, which broke the limit of the Boltzmann kinetic equations (only investigated in the high-temperature domain). Besides, the numerical results show that the AC of a weak EMW in 2D Graphene can have negative values. This demonstrates the possibility of increasing weak EMW by strong laser radiation in 2D Graphene. This is different from a similar problem in bulk semiconductors and the case without strong laser radiation. In the case of  the presence of an external magnetic field, the numerical calculation results also show the appearance of the peak spectral lines that obey the magneto-phonon resonance conditions. The appearance of these resonance peaks provides a model illustrating the dependence of the Half-Width at Half Maximum (HWHM) on the external magnetic field. This is an important criterion for the fabrication of graphene-related electronic components and orientation for future experiments. 
\end{abstract}
%keywords
\begin{keyword}
absorption coefficient \sep 2D graphene \sep magneto-phonon resonance condition \sep strong electromagnetic wave \sep quantum kinetic equation \sep electron-phonon scattering \sep Half-Width at Half Maximum 
\end{keyword}

\end{frontmatter}

\section{Introduction}
Recently, graphene is one of the widely studied two-dimensional semiconductor structures. In graphene, the behavior of electrons is similar to that of relativistic particles of zero mass \cite{no1}. Besides, the wave function and energy spectrum of electrons in graphene are also different from those of conventional semiconductors. This leads to a change of electrical and optical properties in graphene compared with other traditional two-dimensional electron systems. 

In optical properties, the absorption of electromagnetic waves (EMW) is a remarkable effect and has been studied extensively both theoretically and experimentally. The problem of nonlinear absorption of strong EMW in conventional semiconductors has been solved by the author Prof. Nguyen Quang Bau and colleagues by the method of quantum kinetic equations \cite{bau,bau1, bau4, phong,phong1}, showing many new and impressive results that overcome the low-temperature limitation of the method Boltzmann classical kinetic equation. However, from an experimental point of view, it is very difficult to directly observe and measure the absorption coefficient of strong EMW. Therefore, it is common to study the effect of strong EMW on the absorption effect of weak EMW in low-dimensional systems. Theoretically, this problem has been solved in conventional two-dimensional systems such as Quantum Wells \cite{bau1998,nhan1} and Superlattice \cite{bau2002} using different methods. Therefore, in this work, by using the method of quantum kinetic equations, we calculate the AC of a weak EMW in graphene under the influence of a strong EMW. The obtained results will be numerically calculated and graphed with specific parameters to clearly see the dependence of the absorption coefficient on the characteristic quantities of EMW.

\section{Absorption coefficient in the absence of magnetic field}
In this paper, we consider a graphene sheet placed in the x - y plane, with a cross-section $L_x \times L_y$. We consider the electronic state of the $\pi$-bands near a K point, the wave function and energy spectrum of the electron are obtained by solving the Weyl equation for Dirac massless fermions, given by \cite{ando1,ando4,ando5} 
\begin{align}\label{wf0}
\left| {n,{\bf{k}}} \right\rangle  = \Psi \left( {\bf{r}} \right) = \frac{{\exp \left( {i{\bf{k}} \cdot {\bf{r}}} \right)}}{{L\sqrt 2 }}\left( {\begin{array}{*{20}{c}}
n\\
{{e^{i\theta \left( {\bf{k}} \right)}}}
\end{array}} \right)
\end{align}
\begin{align}
    {\varepsilon _{n,{\mathbf{k}}}}\left({\bf{k}} \right)  = n{\gamma }\left| {\bf{k}} \right|
\end{align}
Where $n =  \pm 1$ denotes the conduction and valence bands and 
\begin{align}
\begin{array}{*{20}{c}}
  {{k_x} = \left| {\bf{k}} \right|\cos \theta \left( {\mathbf{k}} \right)}&{{k_y} = \left| {\bf{k}} \right|\sin \theta \left( {\mathbf{k}} \right)} 
\end{array}
\end{align}
$\mathbf{k}$ is the wave vector, ${L_y} = {L_x} = L$ is the sample normalization length in the directions of the system, $\gamma  = 6.46{\text{eV}}{\text{.{\AA}}}$ being the band parameter. 

We consider a field consisting of two planar and monochromatic electromagnetic waves: A strong wave with a frequency $\Omega_1$ and an intensity $\mathbf{E}_{01}$ and a weak wave with frequency $\Omega_2$ and an intensity $\mathbf{E}_{02}$ 
\begin{align}\label{bte}
    {\bf{E}} = {{\bf{E}}_{01}}\sin \left( {{\Omega _1}t + {\varphi _1}} \right) + {{\bf{E}}_{02}}\sin \left( {{\Omega _2}t} \right)
\end{align}
\hl{The corresponding vector potential of this field of two EMWs is given by ${\mathbf{E}} =  - \dfrac{1}{c}\dfrac{{{\text{d}}{\mathbf{A}}}}{{{\text{d}}t}}$ as in previous works} \cite{eps, eps1}. \hl{In these studies, within the framework of dipole approximation, the authors ignored the spatially dependent component of electromagnetic waves. Therefore, the corresponding vector potential of the electromagnetic wave in Eq.} \eqref{bte} \hl{has the form}

\begin{align}%\mathcolorbox{yellow}
    {{\bf{A}} = \frac{c}{\Omega_1}{\bf{E}}_{01}\cos \left( {{\Omega _1}t + {\varphi _1}} \right) +  \frac{c}{\Omega_2}{{\bf{E}}_{02}}\cos \left( {{\Omega _2}t} \right)}
\end{align}

The Hamiltonian of the electron-phonon system in 2D Graphene in the second quantization representation is written as 
\begin{align}
    \begin{split}
         \rm{H} &= \sum\limits_{n,{{\bf{k}}_y}} {{\varepsilon _{n,{{\bf{k}}_y}}}\left( {{{\bf{k}}_y} - \frac{e}{{\hbar c}}{\bf{A}}\left( t \right)} \right)a_{n,{{\bf{k}}_y}}^\dag {a_{n,{{\bf{k}}_y}}}} +  \sum\limits_{\bf{q}} {\hbar {\omega _{\bf{q}}}\left( {b_{\bf{q}}^\dag {b_{\bf{q}}} + \frac{1}{2}} \right)} \\
    &+  \sum\limits_{n,n'} {\sum\limits_{{{\bf{k}}_y},{{\bf{q}}_y}} {{\rm{C}}\left( {\bf{q}} \right)a_{n',{{\bf{k}}_y} + {{\bf{q}}_y}}^\dag {a_{n,{{\bf{k}}_y}}}\left( {b_{ - {\bf{q}}}^\dag  + {b_{\bf{q}}}} \right)} }
    \end{split}
\end{align}
Here, $a_{n,{{\mathbf{k}}_y}}^ \dag $ and ${a_{n,{{\mathbf{k}}_y}}}$ ($ {b_{\mathbf{q}}^ \dag}$  and  ${{b_{\mathbf{q}}}}$) are the creation and annihilation operators of electron (phonon), respectively. 
\hl{In this paper, we limit our calculations to the interaction of electrons with zone-edge optical phonons (mode K), assuming non-dispersive phonons  $\hbar {\omega _{\mathbf{q}}} = \hbar {\omega _{\text{K}}} \equiv \hbar {\omega _0} = 162{\text{meV}}$} \cite{ando2, phuc, h1}. $\rm{C_{\bf{q}}}$ is the electron-optical phonon interaction constant \cite{kry}
\begin{align}\label{cop}
{\left| {{\rm{C}}\left( {\bf{q}} \right)} \right|^2} = \frac{{{\hbar ^2}\rm{D_{op}^2}}}{{2\rho_{g} {L^2}\left( {\hbar {\omega _0}} \right)}}
\end{align}
with $D_{op}$ and $\rho_{g}$ are deformed potential of optical phonon and two-dimensional mass density of graphene.

The absorption coefficient of a weak EMW in the presence of laser radiation in 2D Graphene takes the form \cite{eps}
\begin{align}\label{alp}
    \alpha  = \frac{{8\pi }}{{c\sqrt {{\chi _\infty }} E_{02}^2}}{\left\langle {{{\bf{J}}_y}{{\bf{E}}_{02}}\sin {\Omega _2}t} \right\rangle _t}
\end{align}
where $\chi_\infty$ is the high-frequency dielectric constant in graphene, c is the speed of light in a vacuum, ${\left\langle ...  \right\rangle _t}$ denotes the statistical average value at the moment t. 

In Eq. \eqref{alp}, ${{{\bf{J}}_y}}$ is the current density vector given by \cite{bau1998,bau2002} 
\begin{align}\label{j}
{{\mathbf{J}}_y}\left( t \right) = \frac{{e\hbar }}{{{m_e}}}\sum\limits_{n,{{\mathbf{k}}_y}} {\left( {{{\mathbf{k}}_y} - \frac{e}{{\hbar c}}{\mathbf{A}}\left( t \right)} \right){f_{n,{{\mathbf{k}}_y}}}\left( t \right)} 
\end{align}
Here, $e$ and $m_e$ are the charges and the effective mass of electrons in graphene. ${f_{n,{{\mathbf{k}}_y}}}\left( t \right) = {\left\langle {a_{n,{{\mathbf{k}}_y}}^\dag {a_{n,{{\mathbf{k}}_y}}}} \right\rangle _t}$ is the electron distribution function obtained from solving the general quantum kinetic equation \cite{eps} 

\begin{align}\label{qke}
\frac{{\partial {f_{n,{{\mathbf{k}}_y}}}\left( t \right)}}{{\partial t}} =  - \frac{i}{\hbar }{\left\langle {\left[ {a_{n,{{\mathbf{k}}_y}}^ \dag {a_{n,{{\mathbf{k}}_y}}},{\text{H}}} \right]} \right\rangle _t}
\end{align}

By doing some mathematical manipulations with algebraic operators and replacing the unbalanced distribution function with the balanced distribution function (the first-order tautology approximation method) as in previous papers \cite{nhan1,bau,bau1,phong,phong1}, we obtain the solution of the quantum kinetic equation for the electrons in 2D Graphene 

\begin{align}\label{sol}
\begin{split}
&{f_{n,{{\bf{k}}_y}}}\left( t \right) = {\overline f _{n,{{\bf{k}}_y}}} - \frac{1}{\hbar }\sum\limits_{n',{\bf{q}}}^{} {{{\left| {{\rm{C}}\left( {\bf{q}} \right)} \right|}^2}\sum\limits_{\kappa ,\varsigma ,\rho ,\mu  =  - \infty }^{ + \infty } {{\vartheta _\varsigma }\left[ {\frac{{e{E_{01}}}}{{{\hbar ^2}\Omega _1^2}}\left( {n' - n} \right)} \right]{\vartheta _{\kappa  + \varsigma }}\left[ {\frac{{e{E_{01}}}}{{{\hbar ^2}\Omega _1^2}}\left( {n' - n} \right)} \right]} } \\
 &\times {\vartheta _\mu }\left[ {\frac{{e{E_{02}}}}{{{\hbar ^2}\Omega _2^2}}\left( {n' - n} \right)} \right]{\vartheta _{\rho  + \mu }}\left[ {\frac{{e{E_{02}}}}{{{\hbar ^2}\Omega _2^2}}\left( {n' - n} \right)} \right]\frac{{\exp \left\{ { - i\left[ {\kappa {\Omega _1} + \rho {\Omega _2} + i\delta } \right]t - i\kappa {\varphi _1}} \right\}}}{{\kappa {\Omega _1} + \rho {\Omega _2} + i\delta }}\\
 &\times \left\{ {\frac{{{{\overline f }_{n',{{\bf{k}}_y} - {{\bf{q}}_y}}}{{\overline {\rm{N}} }_{\bf{q}}} - {{\overline f }_{n,{{\bf{k}}_y}}}\left( {{{\overline {\rm{N}} }_{\bf{q}}} + 1} \right)}}{{{\varepsilon _{n,{{\bf{k}}_y}}} - {\varepsilon _{n',{{\bf{k}}_y} - {{\bf{q}}_y}}} - \hbar {\omega _0} - \varsigma \hbar {\Omega _1} - \mu \hbar {\Omega _2} + i\hbar \delta }} + \frac{{{{\overline f }_{n',{{\bf{k}}_y} - {{\bf{q}}_y}}}\left( {{{\overline {\rm{N}} }_{\bf{q}}} + 1} \right) - {{\overline f }_{n,{{\bf{k}}_y}}}{{\overline {\rm{N}} }_{\bf{q}}}}}{{{\varepsilon _{n,{{\bf{k}}_y}}} - {\varepsilon _{n',{{\bf{k}}_y} - {{\bf{q}}_y}}} + \hbar {\omega _0} - \varsigma \hbar {\Omega _1} - \mu \hbar {\Omega _2} + i\hbar \delta }}} \right.\\
 &+ \frac{{{{\overline f }_{n',{{\bf{k}}_y} + {{\bf{q}}_y}}}\left( {{{\overline {\rm{N}} }_{\bf{q}}} + 1} \right) - {{\overline f }_{n,{{\bf{k}}_y}}}{{\overline {\rm{N}} }_{\bf{q}}}}}{{{\varepsilon _{n',{{\bf{k}}_y} + {{\bf{q}}_y}}} - {\varepsilon _{n,{{\bf{k}}_y}}} - \hbar {\omega _0} - \varsigma \hbar {\Omega _1} - \mu \hbar {\Omega _2} + i\hbar \delta }}\left. { + \frac{{{{\overline f }_{n',{{\bf{k}}_y} + {{\bf{q}}_y}}}{{\overline {\rm{N}} }_{\bf{q}}} - {{\overline f }_{n,{{\bf{k}}_y}}}\left( {{{\overline {\rm{N}} }_{\bf{q}}} + 1} \right)}}{{{\varepsilon _{n',{{\bf{k}}_y} + {{\bf{q}}_y}}} - {\varepsilon _{n,{{\bf{k}}_y}}} + \hbar {\omega _0} - \varsigma \hbar {\Omega _1} - \mu \hbar {\Omega _2} + i\hbar \delta }}} \right\}
\end{split}
\end{align}
where ${\vartheta _s}\left( x \right)$ is the $s$th-order Bessel function of the argument $x$, ${\overline f _{n,{{\bf{k}}_y}}} {\left( {{{\overline {\rm{N}} }_{\bf{q}}}} \right)}$ is the balanced distribution function of electrons (phonons), and the quantity $\delta$ is infinitesimal and appears due to the assumption of an adiabatic interaction of the EMW.

Finally, putting Eq. \eqref{sol} in Eq. \eqref{j} and then in Eq. \eqref{alp}, we obtain the general expression for the nonlinear AC of the weak EMW under the influence of a strong EMW in 2D graphene
\begin{align}\label{alpha}
    \begin{split}
&\alpha  = \frac{{{e^2}{n_0}\rm{D_{op}^2}}}{{c\sqrt {{\chi _\infty }}{{\left( {{k_B}T} \right)}^3} {m_e}{\gamma ^3}{\rho _g}{\pi ^2}{{\left( {\hbar {\omega _0}} \right)}^2}\Omega _2^3}}\left[ {\left( {1 + 2\Theta _1^2 - \frac{3}{2}\Theta _1^4} \right)\left( {{\Xi _{0,1}} - {\Xi _{0, - 1}}} \right)} \right.\\
&\left. { + \left( {\Theta _1^2 - \Theta _1^4} \right)\left( {{\Xi _{ - 1,1}} + {\Xi _{1,1}} - {\Xi _{ - 1, - 1}} - {\Xi _{1, - 1}}} \right) + \frac{1}{4}\Theta _1^4\left( {{\Xi _{ - 2,1}} + {\Xi _{2,1}} - {\Xi _{ - 2, - 1}} - {\Xi _{2, - 1}}} \right)} \right]
    \end{split}
\end{align}
\begin{align*}
\begin{array}{*{20}{c}}
{{\Theta _1} = \dfrac{{e{E_{01}}}}{{{\hbar ^2}\Omega _1^2}}}&{{\Xi _{\sigma ,\tau }} = \exp \left( {\dfrac{{\sigma \hbar {\Omega _1} + \tau \hbar {\Omega _2} - \hbar {\omega _0}}}{{2{k_B}T}}} \right)}
\end{array}
\end{align*}
Here, $k_B$ is Boltzmann constant and $n_0$ is the electron density in graphene. 

Eq. \eqref{alpha} is the general expression for the AC of a weak EMW in the presence of external laser radiation in 2D Graphene. The quantities ${{\Xi _{0,1}}, {\Xi _{0, - 1}}}$ are the contribution of the mono-photon absorption and emission processes of the weak EMW. The quantities ${{\Xi _{ - 1,1}}, {\Xi _{1,1}}, {\Xi _{ - 1, - 1}}, {\Xi _{1, - 1}}}$ are the contributions of the processes of absorption and emission of a photon of a weak EMW and a photon of a strong EMW. The quantities ${{\Xi _{ - 2,1}}, {\Xi _{2,1}}, {\Xi _{ - 2, - 1}}, {\Xi _{2, - 1}}}$ are the contributions of the processes of absorption and emission of a photon of a weak EMW and two photons of a strong EMW. In addition, we see that the nonlinear AC of a weak EMW with the electron-optical phonon scattering does not depend on $E_{02}$, only depends on $E_{01}, T, {\hbar {\Omega _1}}$, $ {\hbar {\Omega _2}}$. From expression \eqref{alpha}, when we give $E_{01}=0$, we will get the expression of  the absorption coefficient of a weak EMW in the absence of laser radiation field in 2D graphene with optical electron-phonon scattering. 

\section{Absorption coefficient in the presence of an external magnetic field}
The graphene sheet is now placed under the influence of a strong magnetic field $\mathbf{B}=(0,0,B)$. \hl{The vector potential is chosen according to the Landau gauge of the form $\mathbf{A} ' = \left( {0,Bx,0} \right)$}. In the limit of a strong magnetic field, the energy spectrum of the electron is quantized into Landau levels. The wave function and corresponding Landau spectrum of electrons in 2D Graphene are given by \cite{ando2,ningma,yang} 
 \begin{align}\label{wf}
 \Psi \left( {\mathbf{r}} \right) \equiv \left| {n,{{\mathbf{k}}_y}} \right\rangle  = \frac{{{C_n}}}{{\sqrt {{L_y}} }}{e^{i{k_y}y}}\left[ {\begin{array}{*{20}{c}}
  {{S_n}{\Phi _{\left| n \right| - 1}}\left( {x - X} \right)} \\ 
  {{\Phi _{\left| n \right|}}\left( {x - X} \right)} 
\end{array}} \right]
 \end{align}
\begin{align}\label{el}
{\varepsilon _n} = {S_n}\hbar {\omega _B}\sqrt {\left| n \right|} 
\end{align}
with $S_n = 1$ and $S_n = -1$ stand for the conduction and valence bands; ${C_n} = 1$ when n = 0, ${C_n} = {1 \mathord{\left/
 {\vphantom {1 {\sqrt 2 }}} \right.
 \kern-\nulldelimiterspace} {\sqrt 2 }}$ when $n \ne 0$. In Eq. \eqref{wf} $\Phi _{\left| n \right|}\left( x \right)$ represents harmonic oscillator wavefunctions, centered at $X = {k_y}{l_B^2}$ with $l_B = \sqrt {\hbar /eB} $ is the radius of the ground state electron orbit in the $\left( {x,y} \right)$ plane or the magnetic length. $n = 0, \pm 1, \pm 2,...$ being the Landau Level indices and \hl{the normalized harmonic oscillator function ${\Phi _{\left| n \right|}}\left( x \right) = \dfrac{{{i^{\left| n \right|}}}}{{\sqrt {{2^{\left| n \right|}}\left| n \right|!\sqrt \pi  {l_B}} }}\exp \left[ { - \dfrac{1}{2}{{\left( {\dfrac{x}{{{l_B}}}} \right)}^2}} \right]{H_{\left| n \right|}}\left( {\dfrac{x}{{{l_B}}}} \right)$}, with ${\rm{H}_{\left| n \right|}}\left( {{x \mathord{\left/
 {\vphantom {x l}} \right.
 \kern-\nulldelimiterspace} l_B}} \right)$ is the n-th order Hermite polynomial and $\hbar {\omega _B} = {{\sqrt 2 \gamma } \mathord{\left/
 {\vphantom {{\sqrt 2 \gamma } l_B}} \right.
 \kern-\nulldelimiterspace} l_B}$ is the effective magnetic energy \cite{ando2,ando3}
with $\gamma  = 6,46{\text{eV}}{\text{.}}\mathop {\text{A}}\limits^{\text{o}}$  is the band parameter. 

The Hamiltonian in this case of the electron-phonon system has the form \cite{ando2, phuc}
\begin{align}\label{ha1}
\rm{H} = \sum\limits_{n,{{\mathbf{k}}_y}} {{\varepsilon _n}\left( {{{\mathbf{k}}_y} - \frac{e}{{\hbar c}}{\mathbf{A}}\left( t \right)} \right)a_{n,{{\mathbf{k}}_y}}^ + {a_{n,{{\mathbf{k}}_y}}}} + \sum\limits_{\mathbf{q}} {\hbar {\omega _{\mathbf{q}}}\left( {b_{\mathbf{q}}^ + {b_{\mathbf{q}}} + \frac{1}{2}} \right)} 
+ \sum\limits_{n,n'} {\sum\limits_{{{\mathbf{k}}_y},{\mathbf{q}}} {{\rm{M}_{n.n'}}\left( {\mathbf{q}} \right)a_{n',{{\mathbf{k}}_y} + {{\mathbf{q}}_y}}^ + {a_{n,{{\mathbf{k}}_y}}}\left( {b_{ - {\mathbf{q}}}^ +  + {b_{\mathbf{q}}}} \right)} } 
\end{align}
and the matrix factor $\rm{M}_{n.n'}(\mathbf{q})$  as follows \cite{phuc, ningma}
${\left| {{\rm{M}_{n,n'}}\left( {\mathbf{q}} \right)} \right|^2} = {\left| {\rm{C}\left( {\mathbf{q}} \right)} \right|^2}{\left| {{\rm{J}_{n,n'}}\left( {\mathbf{q}} \right)} \right|^2}$ with $\rm{C}\left( {\mathbf{q}} \right)$ is the electron-optical phonon interaction constant according to Eq. \eqref{cop}; ${{\rm{J}_{n,n'}}\left( {\mathbf{q}} \right)}$ given by 
 \begin{align}
 {\left| {{\rm{J}_{n,n'}}\left( {\mathbf{q}} \right)} \right|^2} = C_n^2C_{n'}^2\frac{{m!}}{{\left( {m + j} \right)!}}{e^{ - u}}{u^j}{\left[ {L_m^j\left( u \right) + {S_n}{S_{n'}}\sqrt {\frac{{m + j}}{m}} L_{m - 1}^j\left( u \right)} \right]^2}
 \end{align}
with ${L_m^j\left( x \right)}$ is the associated Laguerre polynomial, $u = {l_B^2}{q^2}/2,{q^2} = q_x^2 + q_y^2,m = \min \left( {\left| n \right|,\left| {n'} \right|} \right),j = \left| {\left| {n'} \right| - \left| n \right|} \right|$

Doing the same as the case of absence from the field in the previous section, we establish the absorption coefficient for the case of electron-optical phonon interaction by transforming the summations over ${\bf{q}}$ and ${\bf{k}}_y$ to integrals as follows \cite{ningma,h1}
\begin{align}\label{aob}
\begin{split}
\alpha  = {\alpha _0}&\sum\limits_{n',n} {\left[ {{{\overline f }_n}\left( {{{\overline {\rm{N}} }_{\bf{q}}} + 1} \right) - {{\overline f }_{n'}}{{\overline {\rm{N}} }_{\bf{q}}}} \right]\left[ {\left( {{D_{0,1}} - {D_{0, - 1}}} \right) - \frac{1}{2}\left( {{K_{0,1}} - {K_{0, - 1}}} \right) + \frac{3}{{32}}\left( {{Q_{0,1}} - {Q_{0, - 1}}} \right) + } \right.} \\
 &+ \frac{1}{4}\left( {{K_{ - 1,1}} - {K_{ - 1, - 1}} + {K_{1,1}} - {K_{1, - 1}}} \right) - \frac{1}{{16}}\left( {{Q_{ - 1,1}} - {Q_{ - 1, - 1}} + {Q_{1,1}} - {Q_{1, - 1}}} \right) + \\
&\left. { + \frac{1}{{64}}\left( {{Q_{ - 2,1}} - {Q_{ - 2, - 1}} + {Q_{2,1}} - {Q_{2, - 1}}} \right)} \right] \\
{\alpha _0} &= \frac{{4{e^2}\pi {\hbar ^2}{\rm{D}}_{{\rm{op}}}^{\rm{2}}{n_0}}}{{c\sqrt {{\chi _\infty }} \rho \hbar {\omega _0}m_e^2\Omega _2^3\ell _B^2}} \\
{D_{s,m}} &= {I_1}\delta \left( {{\varepsilon _{n'}} - {\varepsilon _n} + \hbar {\omega _0} - s\hbar {\Omega _1} - m\hbar {\Omega _2}} \right) \\
{K_{s,m}} &= a_1^2{I_2}\delta \left( {{\varepsilon _{n'}} - {\varepsilon _n} + \hbar {\omega _0} - s\hbar {\Omega _1} - m\hbar {\Omega _2}} \right) \\
{Q_{s,m}} &= a_1^4{I_3}\delta \left( {{\varepsilon _{n'}} - {\varepsilon _n} + \hbar {\omega _0} - s\hbar {\Omega _1} - m\hbar {\Omega _2}} \right)
\end{split}
\end{align}
Here, ${{{\overline f }_n}}=\overline f \left( {{\varepsilon _n}} \right) = {\left\{ {1 + \exp \left[ {{{\left( {{\varepsilon _n} - {\varepsilon _F}} \right)} \mathord{\left/
 {\vphantom {{\left( {{\varepsilon _n} - {\varepsilon _F}} \right)} {\left( {{k_B}T} \right)}}} \right.
 \kern-\nulldelimiterspace} {\left( {{k_B}T} \right)}}} \right]} \right\}^{ - 1}}$ is the Fermi-Dirac distribution function
for an electron, in which, $\varepsilon_{F}$ is Fermi energy and ${a_1} = {{e{E_{01}}} \mathord{\left/
 {\vphantom {{e{E_{01}}} {\left( {{m_e}\Omega _1^2} \right)}}} \right.
 \kern-\nulldelimiterspace} {\left( {{m_e}\Omega _1^2} \right)}}$. The integrals $I_1, I_2, I_3$ appearing in Eq. \eqref{aob} of the form ${I_\ell } = \int_0^\infty  {{q^{2\ell  + 1}}{{\left| {{{\rm{J}}_{n,n'}}\left( u \right)} \right|}^2}dq} $ with $\ell  = 1,2,3$ are calculated through the integrals in the appendix. $\delta(x)$ is the Delta Dirac function. The delta functions in Eq. \eqref{aob} with an argument of zero represent the law of conservation of energy, also known as the selection rule. This rule allows electrons to move between Landau levels via the absorption of photons of EMWs. However, when the argument is zero, the delta functions might be divergent. To overcome this, we will replace the delta functions with the Lorentzian as follows \cite{van3, h1} $\delta \left( \varepsilon  \right) = \frac{1}{\pi }\frac{\Gamma }{{{\varepsilon ^2} + {\Gamma ^2}}}$ with $\Gamma  = \hbar {\omega _B}\sqrt {{{{\hbar ^2}{\rm{D}}_{{\rm{op}}}^{\rm{2}}} \mathord{\left/
 {\vphantom {{{\hbar ^2}{\rm{D}}_{{\rm{op}}}^{\rm{2}}} {\left( {8\pi \rho {\gamma ^2}\hbar {\omega _0}} \right)}}} \right.
 \kern-\nulldelimiterspace} {\left( {8\pi \rho {\gamma ^2}\hbar {\omega _0}} \right)}}} $ is the level width\cite{phuc}.

The analytic expression \eqref{aob} for the absorption coefficient in the presence of a magnetic field is clearly more complicated than in the absence of a magnetic field. In the following section, we will investigate this difference further by performing numerical calculations and graphing with computer programs and numerical methods.

\section{Numerical results and discussions}
In this section, we perform in-depth numerical computations of the nonlinear AC of a weak EMW in the presence of a laser radiation field in 2D graphene in both the absence and presence of an external magnetic field.  The parameters used in computational calculations are as follow: \cite{ando2,ji,h1,phuc,kuba,kry,nose}  $\gamma = 6.46{\text{eV}}{\text{.{\AA}}}, \rho = 7.{7\times 10^{ - 8}}$ ${\rm{g/c}}{{\rm{m}}^{\rm{2}}}, n_0 = 5 \times 10^{15} \rm{m}^{-2}, {D_{op}} = 1.4 \times {10^9}{\text{eV/cm}}, \hbar {\omega _0} = 162{\text{meV}}, {{\chi _\infty }} = 4$. The value of the Fermi energy level in the Fermi - Dirac distribution function can be approximated between the Landau levels n = 0 and n = 1 for electrons \cite{ji}. In other words, ${\varepsilon_{F}}$ = ${{\hbar {\omega _B}} \mathord{\left/
 {\vphantom {{\hbar {\omega _B}} 2}} \right.
 \kern-\nulldelimiterspace} 2}$, with $\hbar \omega_B$ is the effective magnetic energy from Eq. \eqref{el}. In this paper, we consider the Landau levels from $-2$ to $2$. 

\subsection{In the Absence of an External Magnetic Field}
\begin{figure}
    \centering
    \includegraphics[width=0.75\linewidth]{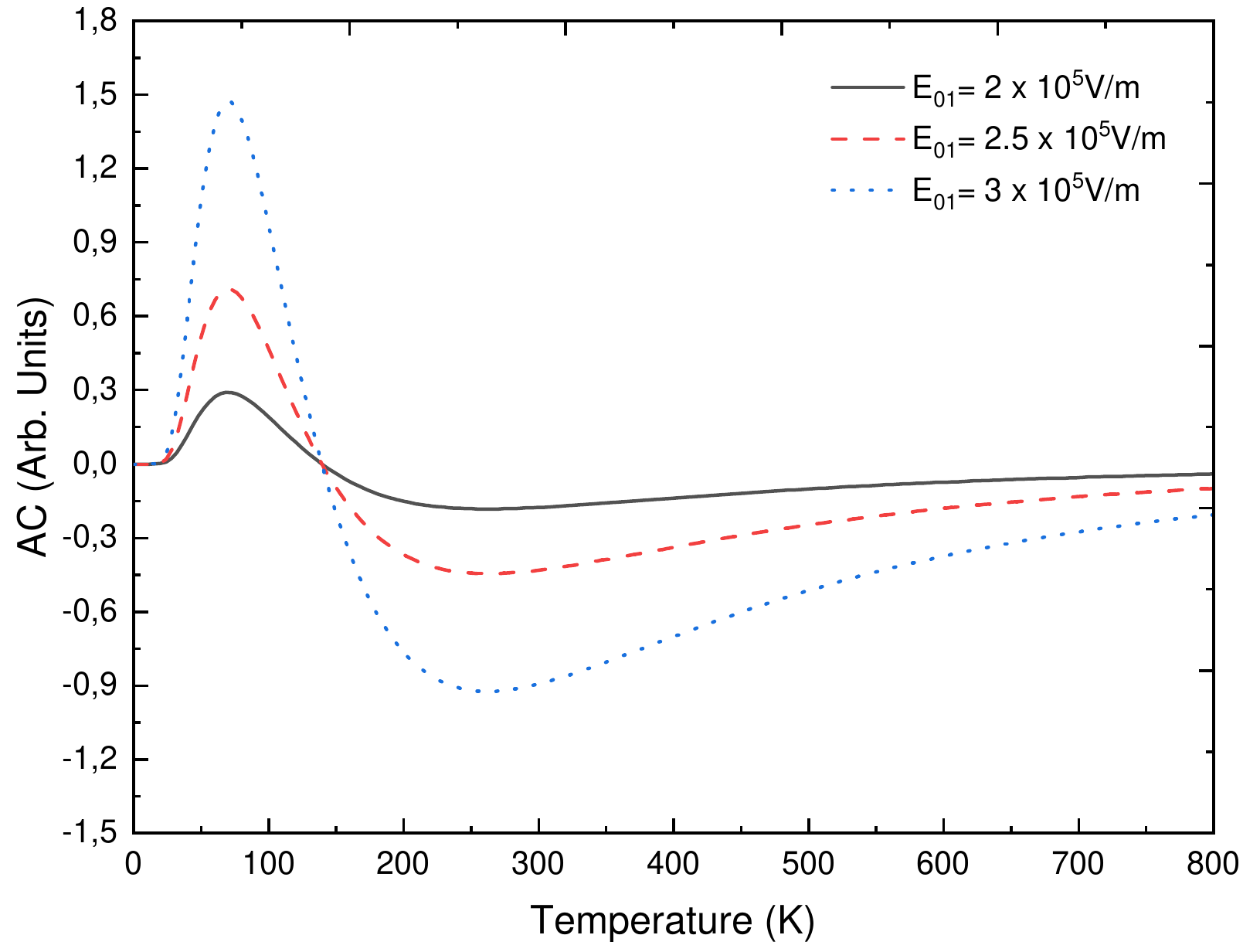}
    \caption{Dependence of AC on the temperature of system. Here, $\hbar {\Omega _2} = 20{\rm{meV}},\hbar {\Omega _1} = 100{\rm{meV}}$}
    \label{alt}
\end{figure}

Figure \ref{alt} describes the dependence of AC on the temperature T for three different values of $E_{01}$. These maximum and minimum values depend on the laser radiation intensity. In addition, we also see that, in Figure \ref{alt}, the AC of a weak EMW in 2D Graphene can have negative values when the temperature is greater than 160K, which means that the weak EMW is increased in the presence of laser radiation. Figure \ref{alt} also shows that in the high-temperature region, the absorption coefficient changes very slowly with temperature and has a small absolute value, i.e. the AC of a weak EMW is small in high-temperature region. This result is similar to previous studies on low-dimensional semiconductors \cite{nhan1,bau2}.

\begin{figure}
\centering
\subfigure[][\label{alw1t}]
  {\includegraphics[width=0.75\linewidth]{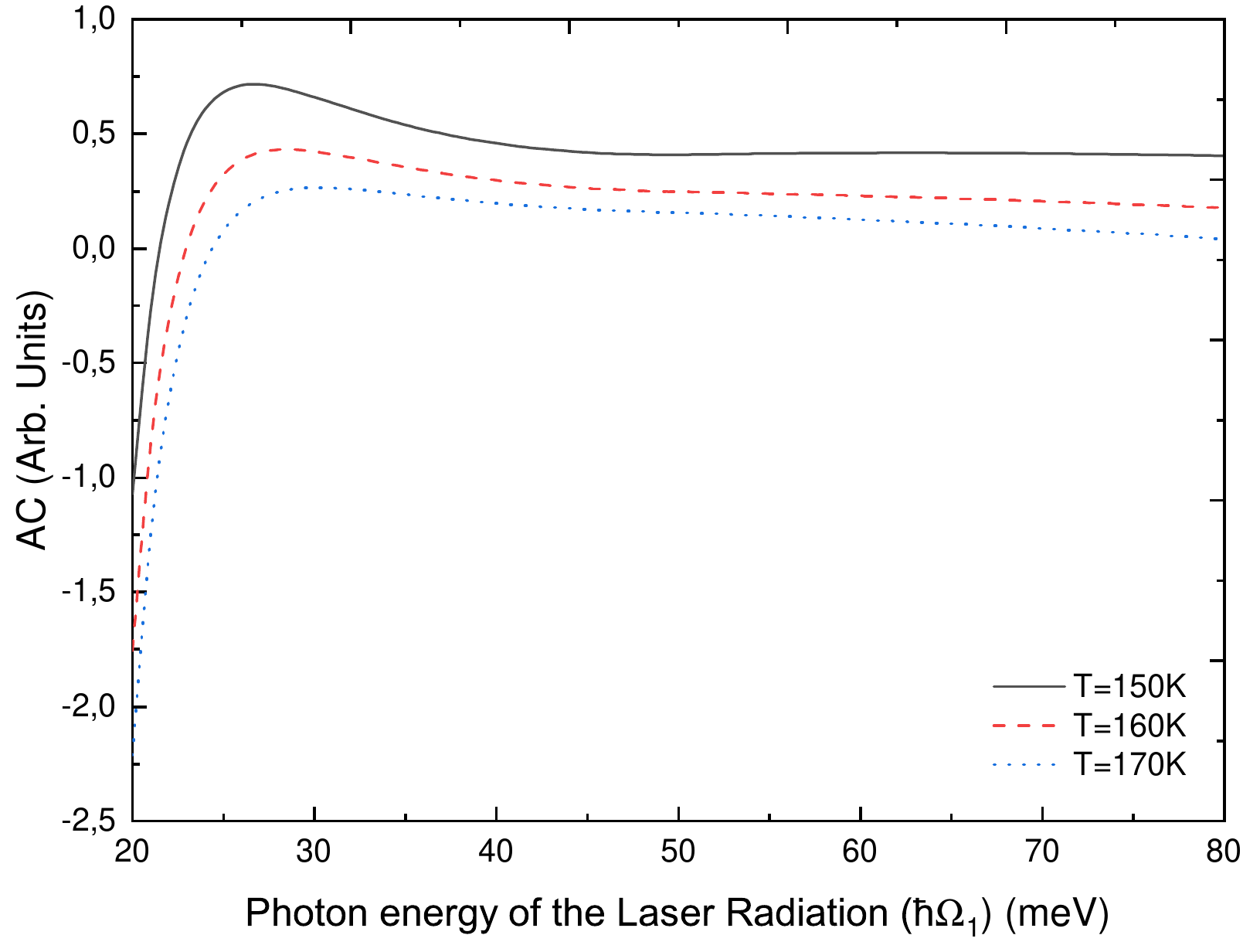}}
\subfigure[][\label{altw1e}]
  {\includegraphics[width=0.75\linewidth]{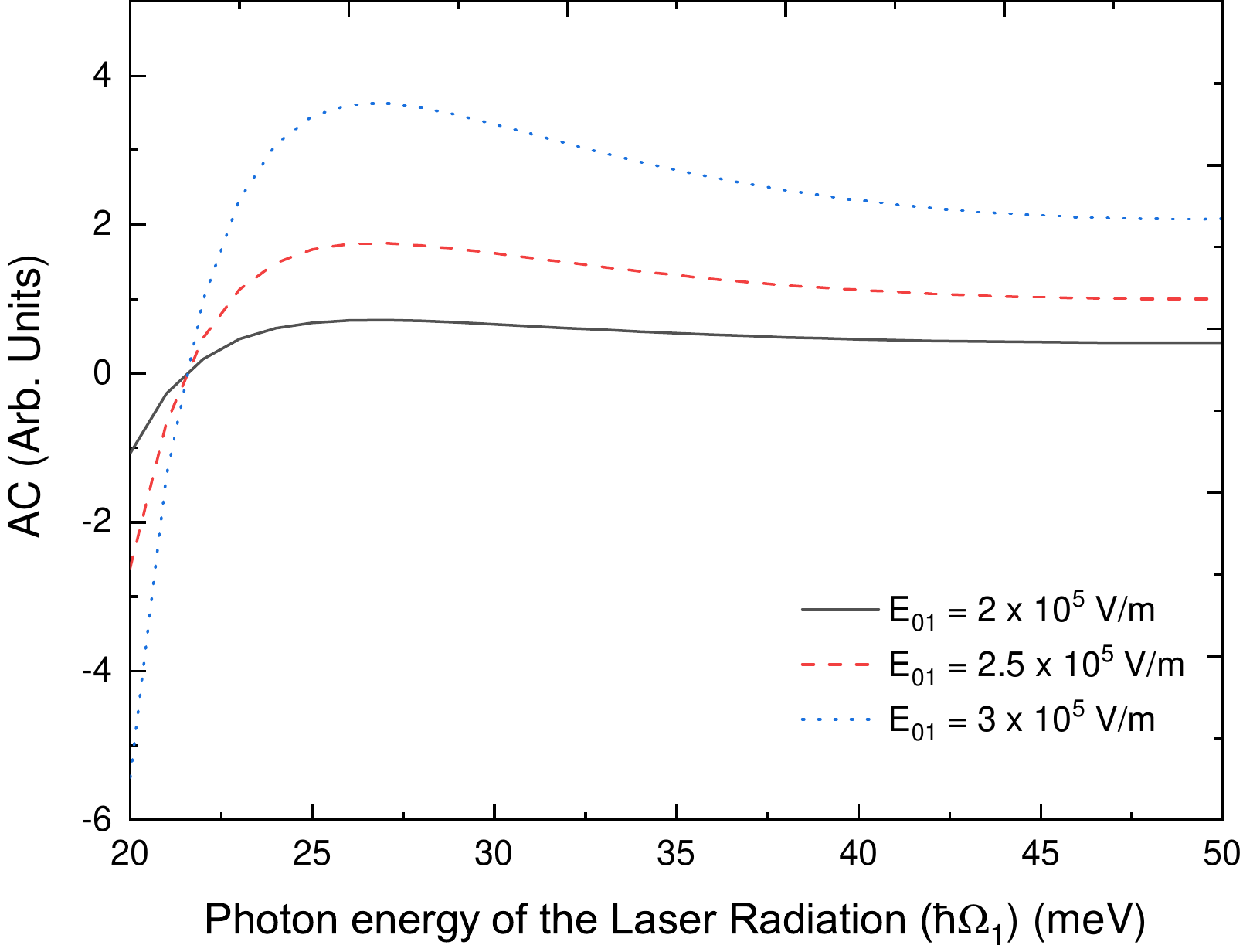}}
\caption{The dependence of AC on the photon energy of the laser Radiation. Here, $\hbar {\Omega _2}=120 \rm{meV}$, (a) $E_{01}=2 \times 10^{5} \rm{V/m}$, (b) T=150K}
\label{altw1}
\end{figure}

In Figure \ref{altw1}, we show the dependence of the AC on the photon energy of the laser Radiation $\hbar {\Omega _1}$. Figure \ref{altw1} shows that when the photon energy of the laser Radiation $\hbar {\Omega _1}$ increases, the AC of a weak EMW also increases to a maximum value, then gradually decreases to a certain value. This absorption peak shifts to the right with increasing $\hbar {\Omega _1}$ as the intensity of the laser radiation $E_{01}$ increases.

\begin{figure}
    \centering
    \includegraphics[width=0.75\linewidth]{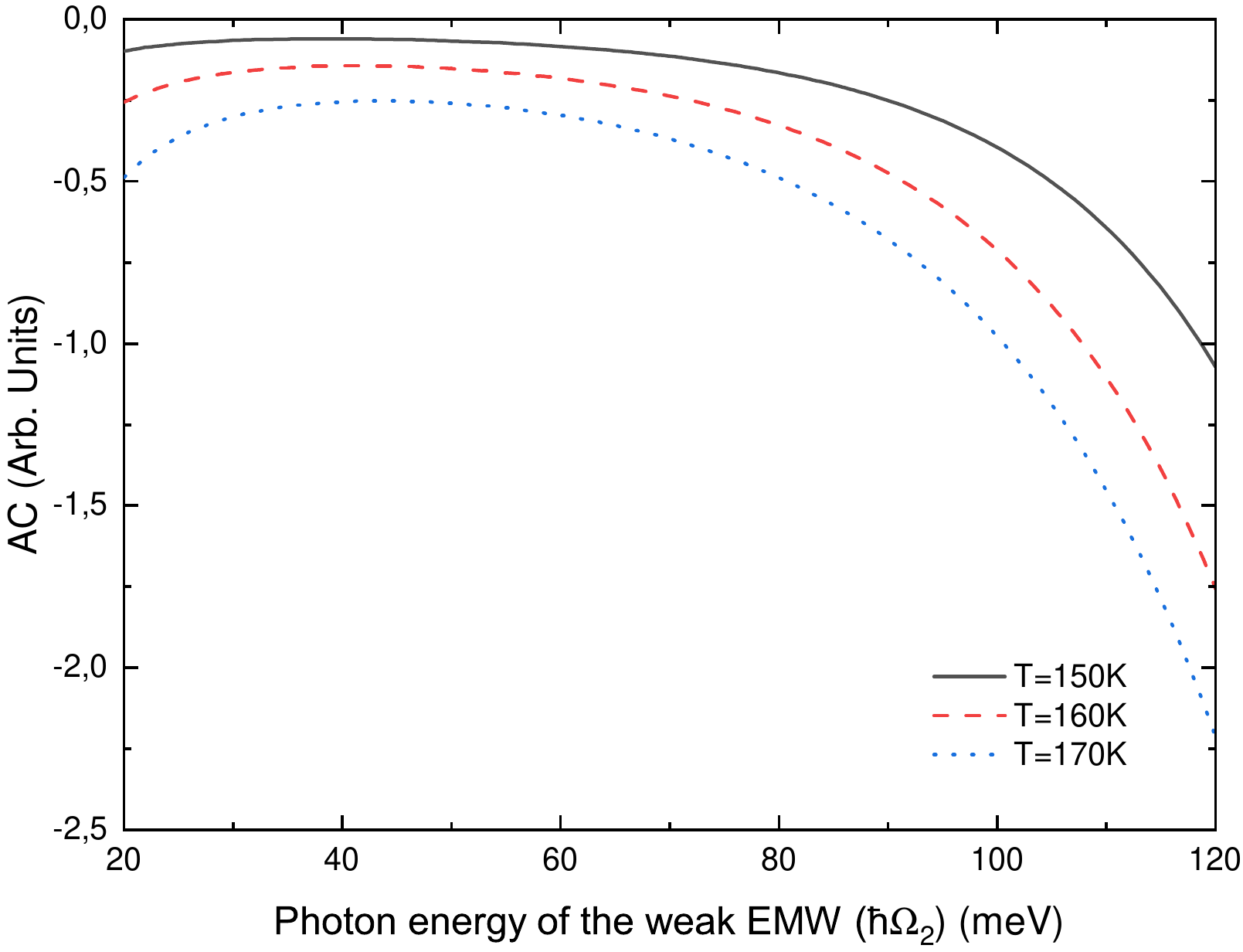}
    \caption{Dependence of AC on the photon energy of the weak EMW. Here, $\hbar {\Omega _1} = 20{\rm{meV}},E_{01}=2 \times 10^{5} \rm{V/m} $}
    \label{alw2}
\end{figure}

Figure \ref{alw2} shows the dependence of AC on the photon energy $\hbar {\Omega _2}$  of the weak EMW at three different values of the temperature of the system. We can again see that AC is negative and the curves have a maximum peak. We can explain this as follows. From figure \ref{alw2} and the analytical expression of AC (Eq. \eqref{alpha}), this maximum peak $\hbar {\Omega _2} \approx 42 \rm{meV}$ satisfies the electron-phonon-photon resonance (EPPR) condition ${{\varepsilon _{n',{{\bf{k}}_y} + {{\bf{q}}_y}}} - {\varepsilon _{n,{{\bf{k}}_y}}} + \hbar {\omega _0} - \varsigma \hbar {\Omega _1} - \mu \hbar {\Omega _2} = 0 }$, when $\varsigma = \mu = 1, k_y \approx  8 \times 10^{7} \rm{m^{-1}}$. In addition, from EPPR condition, we also show that the position of this peak is independent of temperature. 

\begin{figure}
    \centering
    \includegraphics[width=0.75\linewidth]{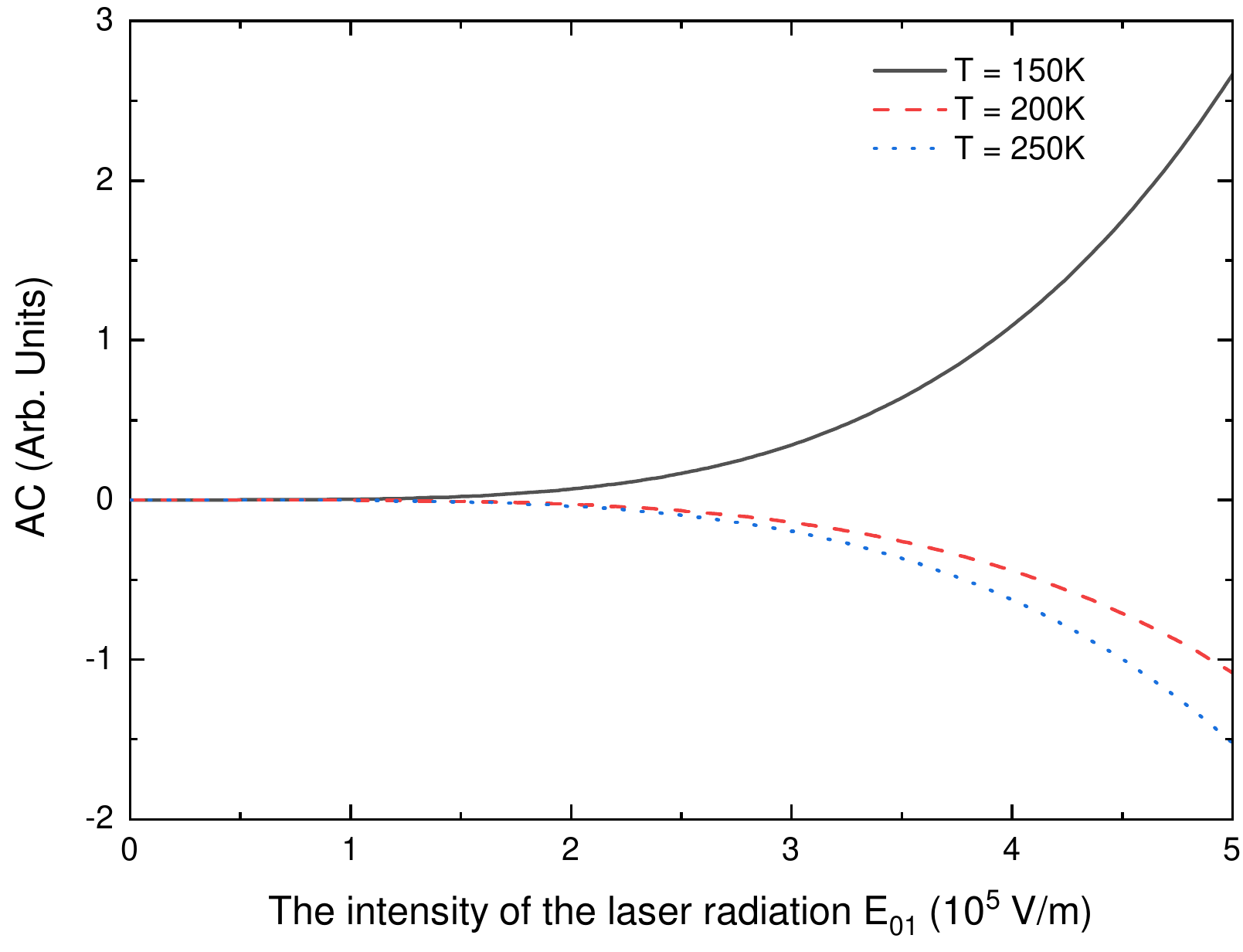}
    \caption{Dependence of AC on the intensity of the Laser Radiation. Here, $\hbar {\Omega _1} = 20{\rm{meV}},\hbar {\Omega _2} = 100{\rm{meV}} $}
    \label{ale11}
\end{figure}

Figure \ref{ale11} shows the dependence of AC on the intensity $E_{01}$ of the Laser Radiation (a strong EMW) at three different values of the temperature of the system. We can see that in the Laser Radiation region of small intensity, the AC is almost unchanged, with the absolute value being very small. At high temperatures, the AC decreases sharply and has a negative value, that is, there is an increase in weak EMW. This can be simply explained as follows. When the temperature is high, the electron mobility increases, leading to an increased scattering probability between the electron and optical phonon, the electron tends to absorb the photon of the strong EMW, giving rise to the weak EMW. 

\subsection{In the Presence of an External Magnetic Field}

\begin{figure}
    \centering
    \includegraphics[width=0.75\linewidth]{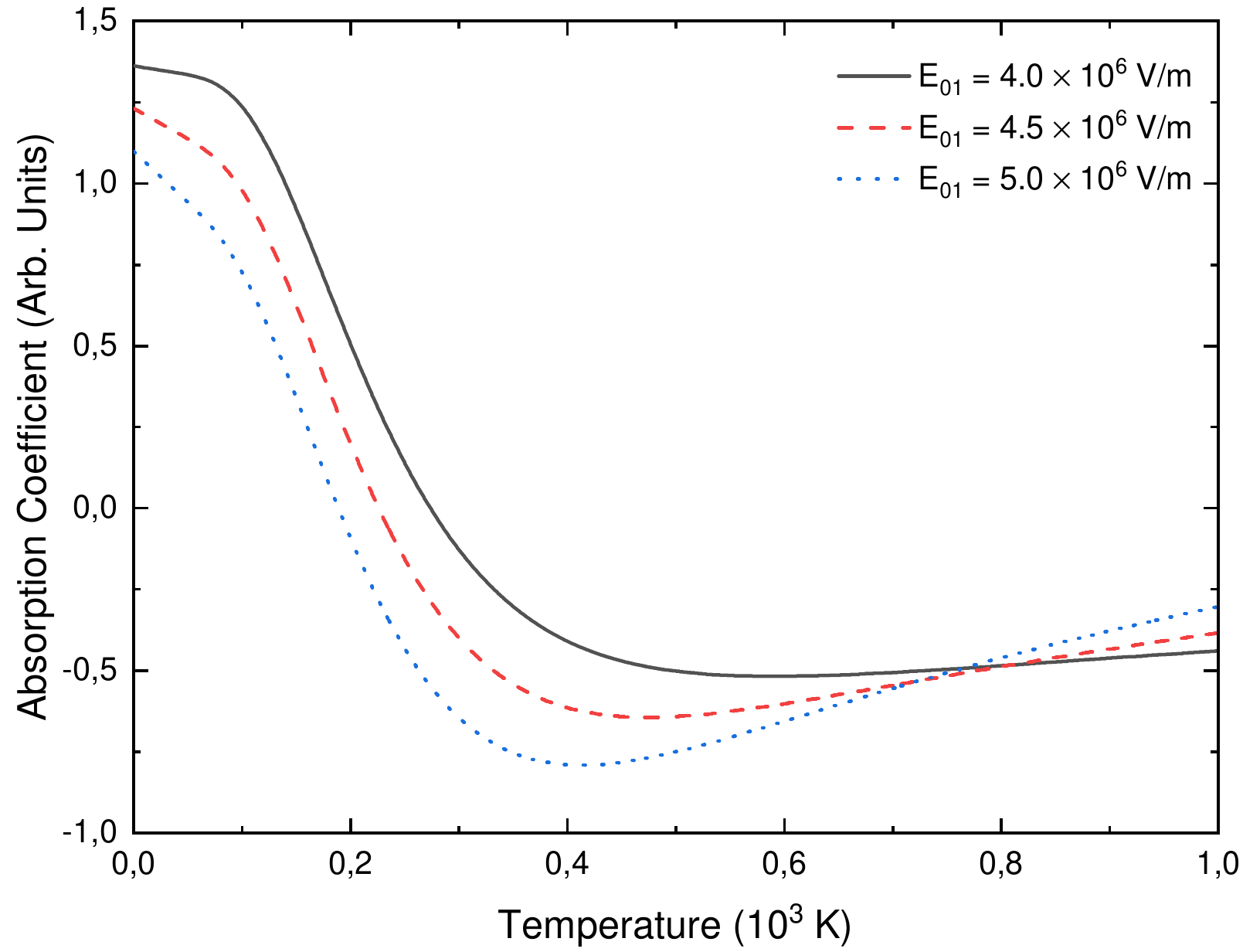}
    \caption{Dependence of AC on the temperature. Here, $\hbar {\Omega _1} = 100{\rm{meV}},\hbar {\Omega _2} = 20{\rm{meV}},B=6T$}
    \label{aOT}
\end{figure}

In Fig. \ref{aOT}, we show the dependence of the AC on the temperature $T$ at three  different values of the intensity of EMW $E_{01}$. Fig. \ref{aOT} shows that when the temperature T of the system rises from 0 K to 1000 K, the curves have a minimum, the value of which depends on the laser radiation intensity. When the temperature is higher, the absorption coefficient changes very slowly and has a very small value. An unsurprising finding is that in the presence of an external magnetic field, the absorption coefficient of weak EMW under the influence of the laser radiation field can still have negative values when the temperature is greater than 280K, this implies that the presence of laser radiation field causes the weak EMW to rise. This is consistent with the result obtained in the case of conventional semiconductors systems \cite{bau2002,bau1998,nhan1} and graphene in the absence of a magnetic field. 

\begin{figure}
    \centering
    \includegraphics[width=0.75\linewidth]{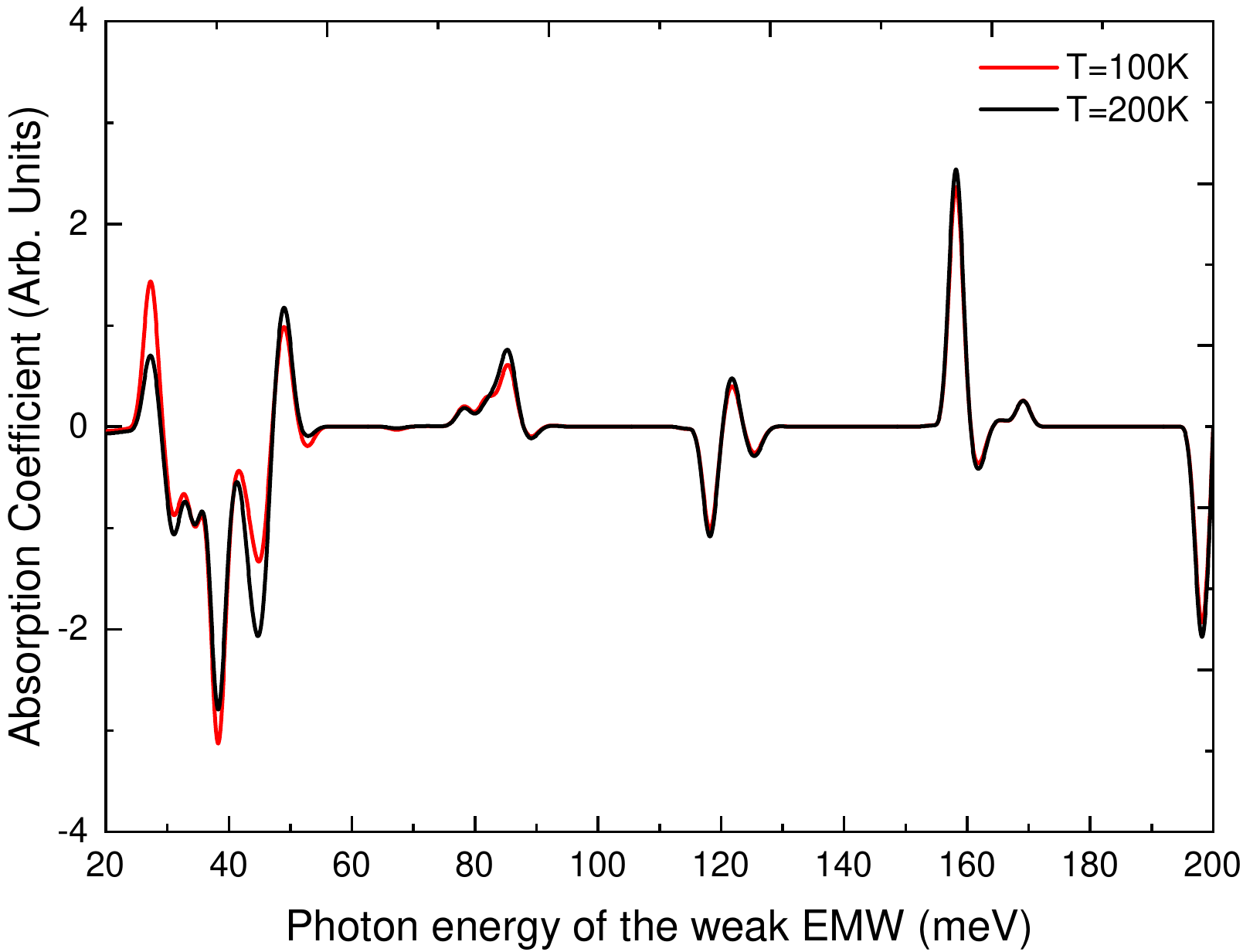}
    \caption{Dependence of AC on the intensity of the photon energy of a weak EMW. Here, $\hbar {\Omega _1} = 40{\rm{meV}},B=6T, E_{01}=5\times10^{6}\rm{V/m}$}
    \label{aO2}
\end{figure}

Plotted in Fig. \ref{aO2} is the AC as a function of the photon energy $\hbar \Omega_2$ of a weak EMW for different values of the temperature T. We can see the appearance of resonance peaks whose position is independent of temperature. The position of the resonance peaks obeys the magneto–phonon-photon resonance condition ${{\varepsilon _{n'}} - {\varepsilon _n} + \hbar {\omega _0} - s\hbar {\Omega _1} - m\hbar {\Omega _2} = 0}$ or $m\hbar {\Omega _2} = \hbar {\omega _B}\left( {{S_{n'}}\sqrt {\left| {n'} \right|}  - {S_n}\sqrt {\left| n \right|} } \right) + \hbar {\omega _0} - s\hbar {\Omega _1}$, where there is a peak occurring at a position where the photon energy of a weak EMW is approximately equal to the optical phonon energy $\hbar {\Omega _2} \approx \hbar {\omega _0} = 162{\rm{meV}}$. This result is a consequence of the law of conservation of energy when the argument of the Delta Dirac functions approaches 0. From the magneto-phonon-photon resonance condition, we also see that the influence of high-frequency electromagnetic waves is significant and causes electron transitions between Landau levels.   

\begin{figure}
    \centering
    \includegraphics[width=0.75\linewidth]{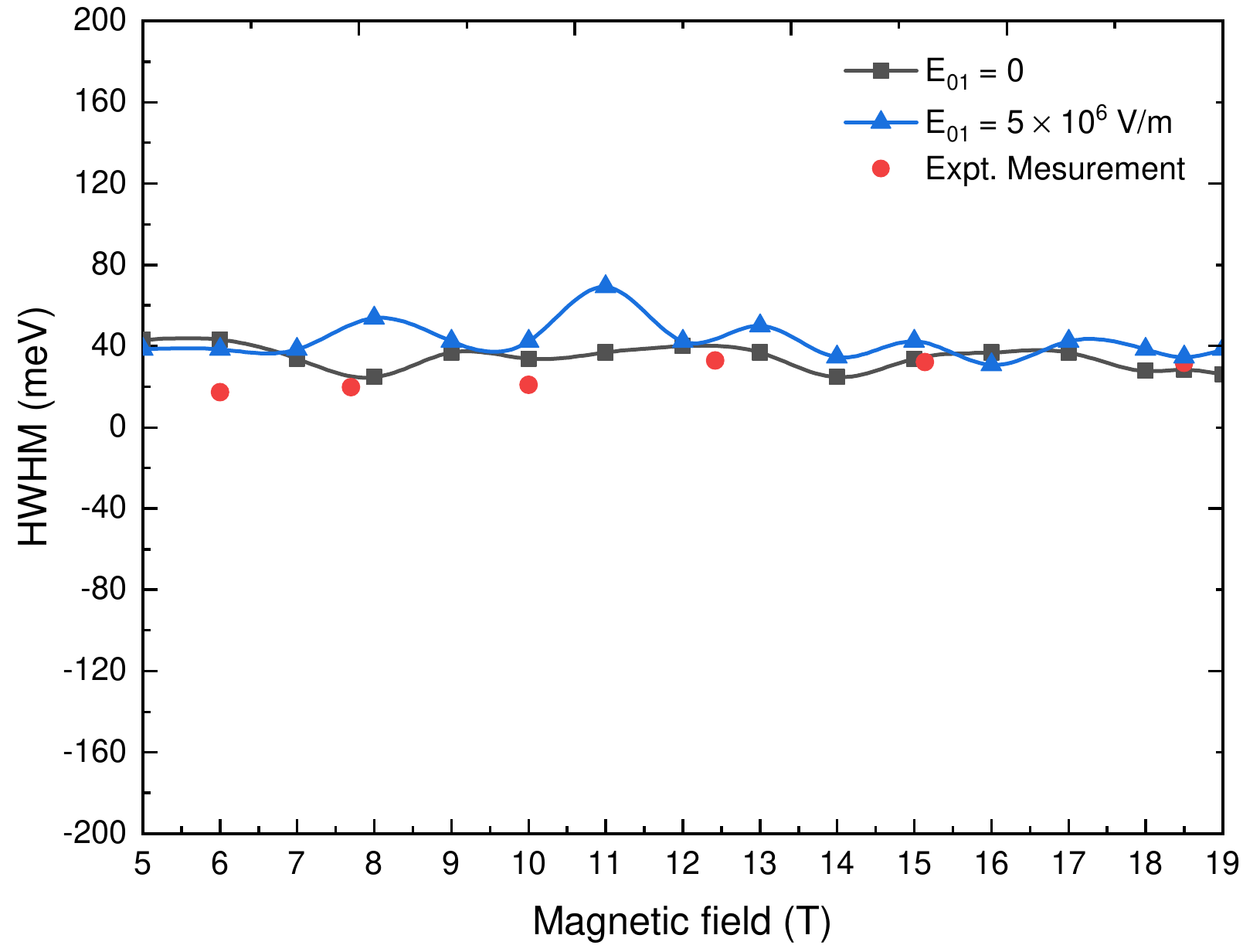}
    \caption{The dependence of the Half-Width at Half Maximum (HWHM) on the magnetic field at T = 4.2K for the transition $n=0$ and $n'=1$. The squares, triangles and circles respectively are our calculations in the case of the system without or under the influence of the laser radiation field and the experimental data taken from ref. \cite{ji}}\label{hwhm}
\end{figure}

Fig. \ref{hwhm} indicates the Half-Width at Half Maximum (HWHM) as a function of the magnetic field. In this figure, we compare the present results to the experimental data acquired earlier by other authors \cite{ji}. Our calculations for zero laser radiation intensity (squares) give similar results to the experimental data (circles). 
We can also see that the HWHM calculation results in the weak magnetic field region are quantitatively different from the experimental data. Indeed, absorption spectral lines appear when there is a transition of electrons between Landau levels that satisfies the magneto-phonon-photon resonance condition. In addition, when the magnetic field is weak, the photon energy of the electromagnetic wave is much stronger than the difference between the two Landau levels of the electron, which leads to our calculated results being larger than the results of experimental observations. In addition, our calculations also show that HWHM increases significantly when placing the system under the influence of a laser radiation field (triangles). This is a new finding that allows us to predict the broadening of the absorption spectrum in the presence of a laser radiation field.

\section{Conclusions}
To sum up, we have solved the problem of nonlinear AC of a weak EMW in 2D graphene in the presence of laser radiation (electron-optical phonon scattering) from the low-temperature to the high-temperature domain. The obtained general analytic expression does not depend on $E_{02}$ but only on $E_{01}, T, \hbar {\Omega _1}$, $\hbar {\Omega _2}$ in both the presence and absence of the external magnetic field. The numerical calculation results of this expression show that the dependence on the above parameters is extremely complicated. Numerical results also show that the value of AC can be negative, which proves that in the presence of a strong EMW, a weak EMW is increased. This is a completely different result from a similar problem in bulk semiconductors. However, this result is completely similar to what has been obtained in previous low-dimensional semiconductor studies. To the best of our knowledge, there are currently no experimental observations that directly test this theoretical prediction. Besides, in the case of absence from the external magnetic field, from the general expressions, the EPPR condition can be obtained by setting arguments in delta functions equal to zero. The maximum peak appears at $\hbar {\Omega _2} \approx 42 \rm{meV}$ satisfying the electron-phonon-photon resonance (EPPR) condition ${{\varepsilon _{n',{{\bf{k}}_y} + {{\bf{q}}_y}}} - {\varepsilon _{n,{{\bf{k}}_y}}} + \hbar {\omega _0} - \varsigma \hbar {\Omega _1} - \mu \hbar {\Omega _2} = 0 }$, when $\varsigma = \mu = 1, k_y \approx  8 \times 10^{7} \rm{m^{-1}}$. In addition, from the EPPR condition, we also show that the position of this peak is independent of temperature. The same is obtained when investigating the influence of the external magnetic field on the absorption coefficient of a weak EMW. However, the peaks, in this case, appear to obey the magneto–phonon resonance condition ${{\varepsilon _{n'}} - {\varepsilon _n} + \hbar {\omega _0} - s\hbar {\Omega _1} - m\hbar {\Omega _2} = 0}$ and their position does not depend on the temperature of the system. And finally, from the investigation of the Half-Width at Half Maximum, we give a theoretical prediction of the absorption line broadening under the influence of the laser radiation field. Our HWHM calculations agree orders of magnitude with current experimental observations. This is an important criterion that can be used in future observations and measurements in the laboratory as well as in research to improve technology for manufacturing electronic components related to graphene.

\section*{Acknowledgement}
This research is financial by Vietnam National University, Hanoi-VNU Strong Research Group, named Method of Quantum Field Theory, and Application of Theoretical Research to Physical Phenomena in quantum environment (lead by Prof. Dr. Nguyen Quang Bau). 

\appendix 
\section{The integrals used in the calculation}
From Eqs. A1, A2, A3 and A4 of Ref. \cite{phuc}, we have 
\begin{align}
{I_1} &= \frac{{m!}}{{\left( {m + j} \right)!}}\int_0^{ + \infty } {{e^{ - u}}{u^{j + 3}}{{\left[ {L_m^j\left( u \right)} \right]}^2}du} \\
 &= \left( {2m + j + 3} \right)\left\{ {2 + 6m\left( {m + 1} \right) + j\left[ {j + 3\left( {2m + 1} \right)} \right]} \right\} + 4m\left( {2m + j} \right)\left( {m + j} \right) \nonumber \\
 {I_2} &= \frac{{\left( {m - 1} \right)!}}{{\left( {m + j} \right)!}}\int_0^{ + \infty } {{e^{ - u}}{u^{j + 3}}L_m^j\left( u \right)L_{m - 1}^j\left( u \right)du}  =  - 3\left( {1 + {j^2} + 5mj + 5{m^2}} \right)
\end{align}

\bibliography{refs}

\begin{thebibliography}{10}
\expandafter\ifx\csname url\endcsname\relax
  \def\url#1{\texttt{#1}}\fi
\expandafter\ifx\csname urlprefix\endcsname\relax\def\urlprefix{URL }\fi
\expandafter\ifx\csname href\endcsname\relax
  \def\href#1#2{#2} \def\path#1{#1}\fi

\bibitem{no1}
K.~S. Novoselov, A.~K. Geim, S.~V. Morozov, D.~Jiang, M.~I. Katsnelson, I.~V. Grigorieva, S.~V. Dubonos, A.~A. Firsov, Two-dimensional gas of massless dirac fermions in graphene, Nature. 438 (2005) 197200.

\bibitem{bau}
N.~Q. Bau, D.~M. Hung, N.~B. Ngoc, The nonlinear absorption coefficient of a strong electromagnetic wave caused by confined electrons in quantum wells, J. Korean Phys. Soc. 54 (2009) 765773.

\bibitem{bau1}
N.~Q. Bau, N.~V. Hieu, N.~V. Nhan, The quantum acoustomagnetoelectric field in a quantum well with a parabolic potential, Superlattices microstruct. 52 (2012) 921930.

\bibitem{bau4}
P.~N. Thang, N.~Q. Bau, et~al., Theoretical study of the influence of confined phonons and a strong electromagnetic wave on the hall effect in an one--dimensional cylindrical quantum wire gaas/gaasal, Mater. Trans. 61~(8) (2020) 1468--1472.

\bibitem{phong}
T.~C. Phong, V.~T. Lam, B.~D. Hoi, Calculation of the nonlinear absorption coefficient of an intense electromagnetic wave caused by confined electrons in doped semiconductor superlattices, J. Korean Phys. Soc. 57 (2010) 12381243.

\bibitem{phong1}
L.~T.~T. Phuong, H.~V. Phuc, T.~C. Phong, Calculation of the nonlinear absorption coefficient of a strong electromagnetic wave by confined electrons in quantum wires, Comput. Mater. Sci 49 (2010) 52605262.

\bibitem{bau1998}
N.~Q. Bau, T.~C. Phong, Calculations of the absorption coefficient of a weak electromagnetic wave by free carriers in quantum wells by the kubo-mori method, Phys. Soc. Jpn. 67~(11) (1998) 3875--3880.

\bibitem{nhan1}
N.~V. Nhan, N.~T.~T. Nhan, N.~V. Nghia, S.~T.~L. Anh, N.~Q. Bau, Ability to increase a weak electromagnetic wave by confined electrons in quantum wells in the presence of laser radiation, in: PIERS Proceedings, Kuala Lumpur, MALAYSIA, 2012, p. 10541059.

\bibitem{bau2002}
N.~Q. Bau, N.~V. Nhan, T.~C. Phong, Calculations of the absorption coefficient of a weak electromagnetic wave by free carriers in doped superlattices by using the kubo-mori method, J. Korean Phys. Soc. 41~(1) (2002) 149--154.

\bibitem{ando1}
T.~Ando, Theory of electronic states and transport in carbon nanotubes, J. Phys. Soc. Jpn. 74 (2005) 777817.

\bibitem{ando4}
T.~Ando, Anomaly of optical phonon in monolayer graphene, J. Phys. Soc. Jpn. 75 (2006) 124701.

\bibitem{ando5}
T.~Ando, The electronic properties of graphene and carbon nanotubes, NPG Asia Mater. 1 (2009) 1721.

\bibitem{eps}
{\'E}.~{\'E}pshtein, Effect of a strong electromagnetic wave on the electronic properties of semiconductors (survey), Radiophysics and Quantum Electronics 18 (1975) 579--598.

\bibitem{eps1}
N.~H. Shon, H.~Nazareno, Propagation of elastic waves in semiconductor superlattices under the action of a laser field, Phys. Rev. B. 50~(3) (1994) 1619.

\bibitem{ando2}
T.~Ando, Magnetic oscillation of optical phonon in graphene, J. Phys. Soc. Jpn. 76 (2007) 024712.

\bibitem{phuc}
H.~V. Phuc, N.~N. Hieu, Nonlinear optical absorption in graphene via two-photon absorption process, Opt. Commun. 344 (2015) 1216.

\bibitem{h1}
B.~D. Hoi, L.~T.~T. Phuong, T.~C. Phong, Magneto-optical absorption and cyclotron-phonon resonance in graphene monolayer, J. Appl. Phys. 123 (2018) 094303.

\bibitem{kry}
S.~V. Kryuchkov, E.~I. Kukhar, D.~V. Zavyalov, Absorption of electromagnetic waves by graphene, Phys. Wave Phenom. 21 (2013) 207213.

\bibitem{ningma}
N.~Ma, S.~Zhang, D.~Liu, Mechanical control over valley magnetotransport in strained graphene, Phys. Lett. A. 380.

\bibitem{yang}
C.~H. Yang, F.~M. Peeters, W.~Xu, Landau-level broadening due to electron-impurity interaction in graphene in strong magnetic fields, Phys. Rev. B. 82 (2010) 075401.

\bibitem{ando3}
N.~Mori, T.~Ando, Magnetophonon resonance monolayer graphene, J. Phys. Soc. Jpn. 80 (2011) 044706.

\bibitem{van3}
M.~P. Chaubey, C.~M. Van~Vliet, Transverse magnetoconductivity of quasi-two-dimensional semiconductor layers in the presence of phonon scattering, Phys. Rev. B. 33 (1986) 5617.

\bibitem{ji}
Z.~Jiang, E.~A. Henriksen, L.~C. Tung, J.~Wang, M.~E. Schwartz, M.~Y. Han, P.~Kim, H.~L. Stormer, Infrared spectroscopy of landau levels of graphene, Phys. Rev. Lett. 98 (2007) 197403.

\bibitem{kuba}
S.~S. Kubakaddi, Interaction of massless dirac electrons with acoustic phonons in graphene at low temperatures, Phys. Rev. B. 79 (2009) 075417.

\bibitem{nose}
A.~H.~C. Neto, F.~Guinea, N.~M.~R. Peres, K.~S. Novoselov, A.~K. Geim, The electronic properties of graphene, Rev. Mod. Phys. 81 (2009) 109.

\bibitem{bau2}
N.~Q. Bau, N.~T.~T. Nhan, N.~V. Nhan, Negative absorption coefficient of a weak electromagnetic wave caused by electrons confined in rectangular quantum wires in the presence of laser radiation, J. Korean Phys. Soc. 64 (2014) 572--578.

\end{thebibliography}


\end{document}